\documentclass{aamas2017}

\usepackage{hyperref}
\usepackage{url}
\usepackage{textcomp}
\usepackage{floatrow}
\usepackage{graphicx}
\usepackage{amsmath}
\usepackage{amsfonts}
\usepackage{todonotes}
\usepackage{subfigure}
\usepackage{epstopdf}
\usepackage{tabulary}
\usepackage{multirow}
\usepackage{color}
\usepackage{hyphenat}
\usepackage{verbatim}
\usepackage{hyperref}

\definecolor{darkgreen}{RGB}{0,125,0}
\newcounter{jzlNoteCounter}
\newcounter{mlNoteCounter}
\newcounter{jmNoteCounter}
\newcounter{vzNoteCounter}
\newcounter{tgNoteCounter}

\newcommand{\bE}{\mathbb{E}}
\newcommand{\bR}{\mathbb{R}}
\newcommand{\cA}{\mathcal{A}}
\newcommand{\cM}{\mathcal{M}}
\newcommand{\cS}{\mathcal{S}}
\newcommand{\cT}{\mathcal{T}}
\newcommand{\cO}{\mathcal{O}}
\newcommand{\cU}{\mathcal{U}}

\newcommand\inlineeqno{\stepcounter{equation}\ (\theequation)}

\newfloatcommand{capbtabbox}{table}[1][\FBwidth]

\graphicspath{ {}}

\title{Multi-agent Reinforcement Learning in\\Sequential Social Dilemmas}

\numberofauthors{5}

\author{
\alignauthor {\normalsize \bf Joel Z. Leibo$^1$}\\
\affaddr{{\small DeepMind, London, UK}}\\
\email{{\small jzl@google.com}}
\alignauthor {\normalsize \bf Vinicius Zambaldi$^1$}\\
\affaddr{{\small DeepMind, London, UK}}\\
\email{{\small vzambaldi@google.com}}
\alignauthor {\normalsize \bf Marc Lanctot}\\
\affaddr{{\small DeepMind, London, UK}}\\
\email{{\small lanctot@google.com}}
\and  
\vspace{0.1cm}
\alignauthor {\normalsize \bf Janusz Marecki}\\
\affaddr{{\small DeepMind, London, UK}}\\
\email{{\small tartel@google.com}}
\alignauthor {\normalsize \bf Thore Graepel}\\
\affaddr{{\small DeepMind, London, UK}}\\
\email{{\small thore@google.com}}
}

\newcommand{\cut}[1]{}

\newcommand{\be}{\begin{equation}}
\newcommand{\ee}{\end{equation}}
\def\bea#1\eea{\begin{align}#1\end{align}}
\def\bean#1\eean{\begin{align*}#1\end{align*}}
\makeatletter
\renewcommand*\env@matrix[1][\arraystretch]{%
  \edef\arraystretch{#1}%
  \hskip -\arraycolsep
  \let\@ifnextchar\new@ifnextchar
  \array{*\c@MaxMatrixCols c}}
\makeatother

\begin{document}

\maketitle

\begin{abstract}
Matrix games like Prisoner's Dilemma have guided research on social dilemmas for decades. However, they necessarily treat the choice to cooperate or defect as an atomic action. In real-world social dilemmas these choices are temporally extended. Cooperativeness is a property that applies to policies, not elementary actions. We introduce sequential social dilemmas that share the mixed incentive structure of matrix game social dilemmas but also require agents to learn policies that implement their strategic intentions. We analyze the dynamics of policies learned by multiple self-interested independent learning agents, each using its own deep Q-network, on two Markov games we introduce here: 1.\ a fruit Gathering game and 2.\ a Wolfpack hunting game. We characterize how learned behavior in each domain changes as a function of environmental factors including resource abundance. Our experiments show how conflict  can emerge from competition over shared resources and shed light on how the sequential nature of real world social dilemmas affects cooperation.
\end{abstract}

\ccsdesc[500]{Computing methodologies~Multi-agent reinforcement learning}
\ccsdesc[300]{Computing methodologies~Agent / discrete models}
\ccsdesc[300]{Computing methodologies~Stochastic games}

\printccsdesc

\keywords{Social dilemmas, cooperation, Markov games, agent-based social simulation, non-cooperative games}

\section{Introduction}\label{sec:intro}

\footnotetext[1]{These authors contributed equally.}

Social dilemmas expose tensions between collective and individual rationality~\cite{rapoport1974prisoner}. Cooperation makes possible better outcomes for all than any could obtain on their own. However, the lure of free riding and other such parasitic strategies implies a tragedy of the commons that threatens the stability of any cooperative venture \cite{van2013psychology}. 

The theory of repeated general-sum matrix games provides a framework for understanding social dilemmas. Fig. \ref{fig:matrix_games} shows payoff matrices for three canonical examples: Prisoner's Dilemma, Chicken, and Stag Hunt. The two actions are interpreted as cooperate and defect respectively. The four possible outcomes of each stage game are $R$ (reward of mutual cooperation), $P$ (punishment arising from mutual defection), $S$ (sucker outcome obtained by the player who cooperates with a defecting partner), and $T$ (temptation outcome achieved by defecting against a cooperator). A matrix game is a social dilemma when its four payoffs satisfy the following \emph{social dilemma inequalities} (this formulation from \cite{macy2002learning}):

\begin{enumerate}
 \item $R > P$  ~~~~~~~~ Mutual cooperation is preferred to mutual defection. \hfill $\label{eq:RgeP} \inlineeqno$
 \item $R > S$  ~~~~~~~~ Mutual cooperation is preferred to being exploited by a defector. \hfill $\label{eq:RgeS} \inlineeqno$
 \item $2R > T + S$ ~ This ensures that mutual cooperation is preferred to an equal probability of unilateral cooperation and defection. \hfill $\label{eq:noMixing} \inlineeqno$
 \item either \emph{greed}: $T > R$  ~~~ Exploiting a cooperator is preferred over mutual cooperation \\
       or ~~~ \emph{fear}:  $P > S$  ~~~~ Mutual defection is preferred over being exploited. \hfill $\label{eq:greedOrfear} \inlineeqno$
\end{enumerate}

Matrix Game Social Dilemmas (MGSD) have been fruitfully employed as models for a wide variety of phenomena in theoretical social science and biology. For example, there is a large and interesting literature concerned with mechanisms through which the socially preferred outcome of mutual cooperation can be stabilized, e.g., direct reciprocity~\cite{trivers1971evolution, Axelrod84, nowak1992tit, nowak1993strategy}, indirect reciprocity~\cite{nowak1998evolution}, norm enforcement~\cite{axelrod1986evolutionary, mahmoud2016cooperation}, simple reinforcement learning variants~\cite{macy2002learning}, multiagent reinforcement learning~\cite{sandholm96,MunozDeCote06,wunder10,Zawadzki14,Bloembergen15}, spatial structure~\cite{nowak1992evolutionary}, emotions~\cite{yu2015emotional}, and social network effects~\cite{ohtsuki2006simple, santos2006new}.

\begin{figure*}[t]
\centering
{\normalsize
\begin{tabular}{c|c|c|c|}
  & C & D \\
\hline
C & $R, R$ & $S, T$ \\
\hline
D & $T, S$ & $P, P$ \\
\hline
\end{tabular}
~~ \vline ~~~
\begin{tabular}{c|c|c|c|}
{\small Chicken}  & C & D \\
\hline
C & $3, 3$ & $1, 4$ \\
\hline
D & $4, 1$ & $0, 0$ \\
\hline
\end{tabular}
~~~~~
\begin{tabular}{c|c|c|c|}
 {\small Stag Hunt} & C & D \\
\hline
C & $4, 4$ & $0, 3$ \\
\hline
D & $3, 0$ & $1, 1$ \\
\hline
\end{tabular}
~~~~~
\begin{tabular}{c|c|c|c|}
 {\small Prisoners} & C & D \\
\hline
C & $3 ,3$ & $0, 4$ \\
\hline
D & $4, 0$ & $1, 1$ \\
\hline
\end{tabular}
}
\caption{Canonical matrix game social dilemmas. Left: Outcome variables $R$, $P$, $S$, and $T$ are mapped to cells of the game matrix. Right: The three canonical matrix game social dilemmas.
By convention, a cell of $X,Y$ represents a utility of $X$ to the row player and $Y$ to the column player.
In Chicken, agents may defect out of greed. In Stag Hunt, agents may defect out of fear of a non-cooperative partner. In Prisoner's Dilemma, agents are motivated to defect out of both greed and fear simultaneously.
\label{fig:matrix_games}}
\end{figure*}

However, the MGSD formalism ignores several aspects of real world social dilemmas which may be of critical importance.
\begin{enumerate}
 \item Real world social dilemmas are temporally extended.
 \item Cooperation and defection are labels that apply to \emph{policies} implementing strategic decisions.
 \item Cooperativeness may be a graded quantity.
 \item Decisions to cooperate or defect occur only quasi\hyp{}simultaneously since some information about what player 2 is starting to do can inform player 1's decision and vice versa.
 \item Decisions must be made despite only having partial information about the state of the world and the activities of the other players.
\end{enumerate}

We propose a \emph{Sequential} Social Dilemma (SSD) model to better capture the above points while, critically, maintaining the mixed motivation structure of MGSDs. That is, analogous inequalities to \eqref{eq:RgeP} -- \eqref{eq:greedOrfear} determine when a temporally-extended Markov game is an SSD.

To demonstrate the importance of capturing sequential structure in social dilemma modeling, we present empirical game-theoretic analyses  \cite{walsh2002analyzing, Wellman06} of SSDs to identify the empirical payoff matrices summarizing the outcomes that would arise if cooperate and defect \emph{policies} were selected as one-shot decisions. The empirical payoff matrices are themselves valid matrix games. Our main result is that both of the SSDs we considered, Gathering and Wolfpack, have empirical payoff matrices that are Prisoner's Dilemma (PD). This means that if one were to adhere strictly to the MGSD-modeling paradigm, PD models should be proposed for both situations. Thus any conclusions reached from simulating them would necessarily be quite similar in both cases (and to other studies of iterated PD). However, when viewed as SSDs, the formal equivalence of Gathering and Wolfpack disappears. They are clearly different games. In fact, there are simple experimental manipulations that, when applied to Gathering and Wolfpack, yield \emph{opposite} predictions concerning the emergence and stability of cooperation.

More specifically, we describe a factor that promotes the emergence of cooperation in Gathering while discouraging its emergence in Wolfpack, and vice versa. The straightforward implication is that, for modeling real-world social dilemmas with SSDs, the choice of whether to use a Gathering-like or Wolfpack-like model is critical. And the differences between the two cannot be captured by MGSD modeling.

Along the way to these results, the present paper also makes a small methodological contribution. Owing to the greater complexity arising from their sequential structure, it is more computationally demanding to find equilibria of SSD models than it is for MGSD models. Thus the standard evolution and learning approaches to simulating MGSDs cannot be applied to SSDs. Instead, more sophisticated multiagent reinforcement learning methods must be used (e.g \cite{Littman94markovgames, Nowe12, kleimanWeiner2016}). In this paper we describe how deep Q-networks (e.g \cite{Mnih:2015}) may be applied to this problem of finding equilibria of SSDs.

\begin{figure}
\centering
\includegraphics[width=0.9\linewidth]{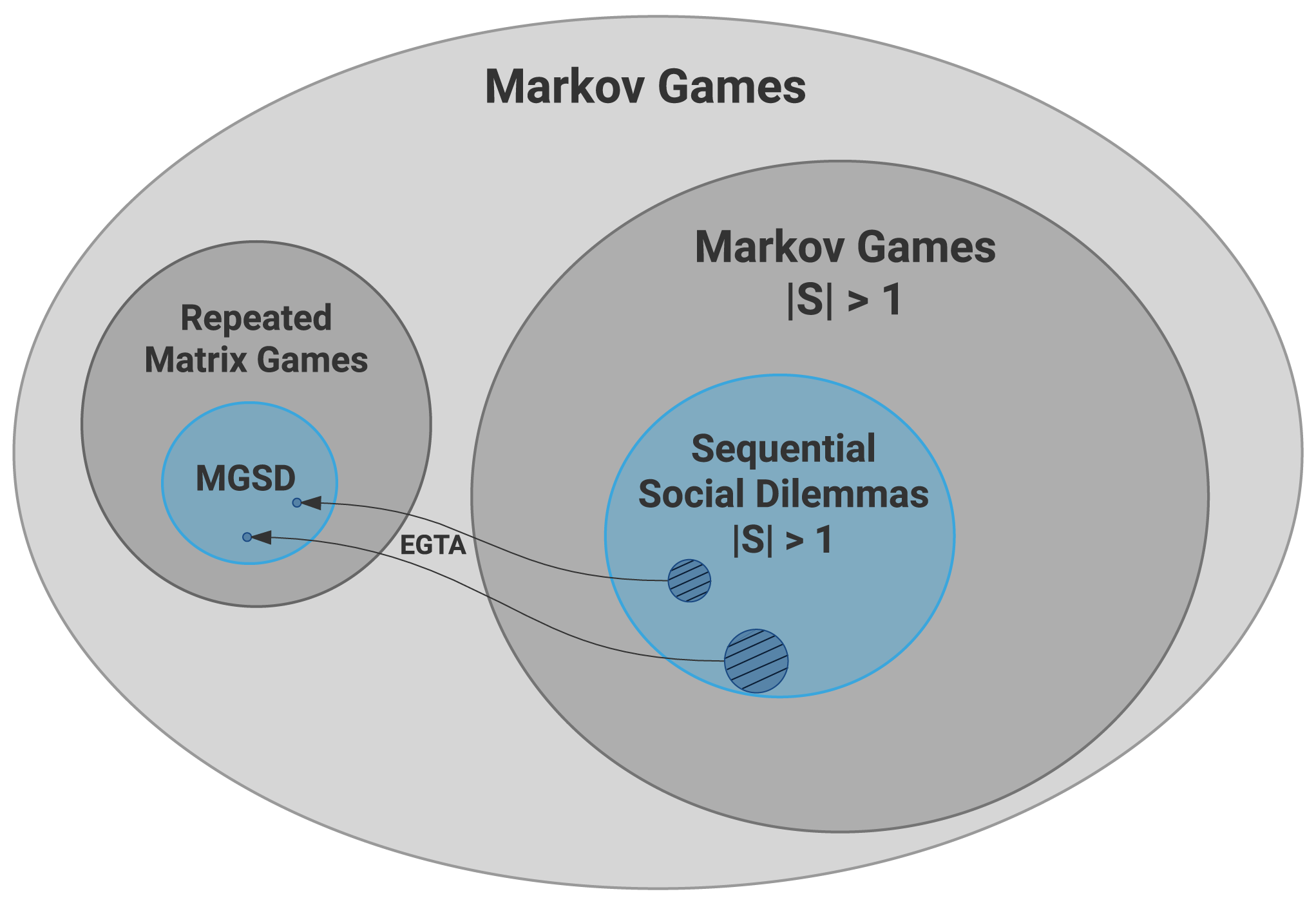}
\caption{Venn diagram showing the relationship between Markov games, repeated matrix games,
MGSDs, and SSDs. A repeated matrix game is an MGSD when it satisfies the social dilemma 
inequalities (eqs. \ref{eq:RgeP} -- \ref{eq:greedOrfear}). A Markov game with $|\cS|>1$
is an SSD when it can be mapped by empirical game-theoretic analysis
(EGTA)
to an MGSD. Many SSDs may map to the same MGSD.}
\label{fig:ssds_venn}
\end{figure}

\vskip 1cm

\section{Definitions and Notation} \label{sec:notation}
We model sequential social dilemmas as general-sum Markov (simultaneous move) games with each agent having only a partial observation onto their local environment. Agents must learn an appropriate policy while coexisting with one another. A policy is considered to implement cooperation or defection by properties of the realizations it generates. A  Markov game is an SSD if and only if it contains outcomes arising from cooperation and defection policies that satisfy the same inequalities used to define MGSDs (eqs. \ref{eq:RgeP} -- \ref{eq:greedOrfear}). This definition is stated more formally in sections \ref{sec:markov_games} and \ref{sec:define_ssd} below.

\subsection{Markov Games} \label{sec:markov_games}

A two-player partially observable Markov game $\cM$ is defined by a set of states $\cS$ and an observation function $O: \cS \times \{1,2\} \rightarrow \bR^d$ specifying each player's $d$-dimensional view, along with two sets of actions allowable from any state $\cA_1$ and $\cA_2$, one for each player, a transition function $\cT: \cS \times \cA_1 \times \cA_2 \rightarrow \Delta(\cS)$, where $\Delta(\cS)$ denotes the set of discrete probability distributions over $\cS$, and a reward function for each player: $r_i: \cS \times \cA_1 \times \cA_2 \rightarrow \bR$ for player $i$.
Let $\cO_i = \{ o_i~|~s \in \cS, o_i = O(s,i)\}$ be the observation space of player $i$.
To choose actions, each player uses policy $\pi_i : \cO_i \rightarrow \Delta(\cA_i)$.

For temporal discount factor $\gamma \in [0, 1]$ we can define the long-term payoff $V^{\vec{\pi}}_i(s_0)$ to player $i$ when the joint policy $\vec{\pi} = (\pi_1, \pi_2)$ is followed starting from state $s_0 \in \cS$. 

\begin{equation}\label{eq:expected_payoff}
V^{\vec{\pi}}_i(s_0) = \bE_{\vec{a}_t \sim \vec{\pi}(O(s_t)), s_{t+1} \sim \cT(s_t, \vec{a}_t)}\left[ \sum_{t = 0}^\infty \gamma^t r_i(s_t, \vec{a}_t) \right].
\end{equation}

Matrix games are the special case of two-player perfectly observable ($O_i(s) = s$) Markov games obtained when $|\cS| = 1$. MGSDs also specify $\cA_1 = \cA_2 = \{C, D\}$, where $C$ and $D$ are called (atomic) cooperate and defect respectively.

The outcomes $R(s), P(s), S(s), T(s)$ that determine when a matrix game is a social dilemma are defined as follows.
\begin{eqnarray}
 &R(s) := V_1^{\pi^C, \pi^C}(s) = V_2^{\pi^C, \pi^C}(s), \label{eq:R}\\
 &P(s) := V_1^{\pi^D, \pi^D}(s) = V_2^{\pi^D, \pi^D}(s), \label{eq:P}\\
 &S(s) := V_1^{\pi^C, \pi^D}(s) = V_2^{\pi^D, \pi^C}(s), \label{eq:S}\\
 &T(s) := V_1^{\pi^D, \pi^C}(s) = V_2^{\pi^C, \pi^D}(s), \label{eq:T}
\end{eqnarray}

where $\pi^C$ and $\pi^D$ are cooperative and defecting {\it policies} as described next.
Note that a matrix game is a social dilemma when $R, P, S, T$ satisfy the inequalities \eqref{eq:RgeP} -- \eqref{eq:greedOrfear}.\\

\subsection{Definition of Sequential Social Dilemma} \label{sec:define_ssd}
This definition is based on a formalization of empirical game-theoretic analysis~\cite{walsh2002analyzing, Wellman06}. We define the  outcomes\newline $(R, P, S, T):=$ $(R(s_0)$, $P(s_0)$, $S(s_0)$, $T(s_0))$ induced by initial state $s_0$, and two policies  $\pi^C, \pi^D$, through their long-term expected payoff \eqref{eq:expected_payoff} and the definitions \eqref{eq:R} -- \eqref{eq:T}. We refer to the game matrix with $R$, $P$, $S$, $T$ organized as in Fig. \ref{fig:matrix_games}-left. as an {\it empirical payoff matrix} following the terminology of \cite{Wellman06}.

\textbf{Definition: } A sequential social dilemma is a tuple $(\cM, \Pi^C, \Pi^D)$ where $\Pi^C$ and $\Pi^D$ are disjoint sets of policies that are said to implement cooperation and defection respectively. $\cM$ is a Markov game with state space $\cS$. Let the empirical payoff matrix $(R(s), P(s), S(s), T(s))$ be induced by policies $(\pi^C \in \Pi^C, \pi^D \in \Pi^D)$ via eqs. \eqref{eq:expected_payoff} -- \eqref{eq:T}. A Markov game is an SSD when there exist states $s \in \cS$ for which the induced empirical payoff matrix satisfies the social dilemma inequalities \eqref{eq:RgeP} -- \eqref{eq:greedOrfear}.

\textbf{Remark: } There is no guarantee that $\Pi^C \bigcup \Pi^D = \Pi$, the set of all legal policies. This reflects the fact that, in practice for sequential behavior, cooperativeness is usually a graded property. Thus we are forced to define $\Pi^C$ and $\Pi^D$ by thresholding a continuous \emph{social behavior metric}. For example, to construct an SSD for which a policy's level of aggressiveness $\alpha:\Pi \rightarrow \mathbb{R}$ is the relevant social behavior metric, we pick threshold values $\alpha_c$ and $\alpha_d$ so that $\alpha(\pi) < \alpha_c \iff \pi \in \Pi^C$ and $\alpha(\pi) > \alpha_d \iff \pi \in \Pi^D$.\\

\begin{figure}
\centering
\begin{minipage}{.5\textwidth}
  \centering
  \includegraphics[width=1\linewidth]{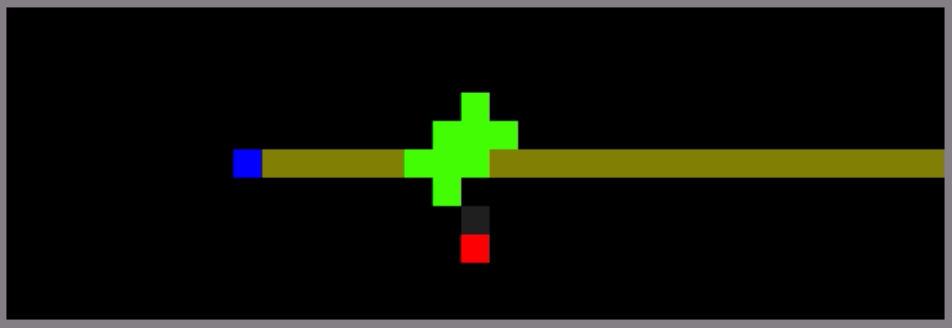}
\end{minipage}%
\begin{minipage}{.5\textwidth}
  \centering
  \includegraphics[width=1\linewidth]{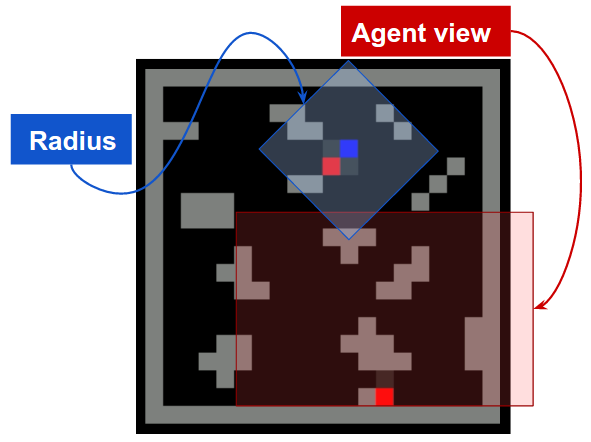}
\end{minipage}
\caption{ Left: Gathering. In this frame the blue player is directing its beam at the apple respawn location. The red player is approaching the apples from the south. Right: Wolfpack. The size of the agent's view relative to the size of the map is illustrated. If an agent is inside the blue diamond-shaped region around the prey when a capture occurs---when one agent touches the prey---both it and its partner receive a reward of $r_{\text{team}}$.} 
\label{fig:maps}
\end{figure}

\section{Learning Algorithms} \label{sec:indep_deep_learning}

Most previous work on finding policies for Markov games takes the prescriptive view of multiagent learning~\cite{Shoham07}: that is, it attempts to answer ``what {\it should} each agent do?'' Several algorithms and analyses have been developed for the two-player zero-sum case~\cite{Littman94markovgames,Lagoudakis02,Perolat15,Perolat16Softened,Bosansky16Algorithms}. The general-sum case is significantly more challenging~\cite{Zinkevich06}, and algorithms either have strong assumptions or need to either track several different potential equilibria per agent~\cite{Hu98NashQ,Greenwald03CEQ}, model other players to simplify the problem~\cite{Littman01}, or must find a cyclic strategy composed of several policies obtained through multiple state space sweeps~\cite{Perolat16}. Researchers have also studied the emergence of multi-agent coordination in the decentralized, partially observable MDP framework \cite{gmytrasiewicz2005framework, varakantham2009exploiting, becker2004solving}. However, that approach relies on knowledge of the underlying Markov model, an unrealistic assumption for modeling real-world social dilemmas. 

In contrast, we take a descriptive view, and aim to answer ``what social effects emerge when each agent uses a particular learning rule?'' The purpose here then is to study and characterize the resulting learning dynamics, as in e.g.,~\cite{wunder10, Bloembergen15}, rather than on designing new learning algorithms. It is well-known that the resulting ``local decision process'' could be non-Markovian from each agent's perspective~\cite{Laurent11}. This is a feature, not a bug in descriptive work since it is a property of the real environment that the model captures. 

We use deep reinforcement learning as the basis for each agent in part because of its recent success with solving complex problems~\cite{Mnih:2015,Silver16Go}. Also, temporal difference predictions have been observed in the brain~\cite{Schultz97} and this class of reinforcement learning algorithm is seen as a candidate theory of animal habit-learning~\cite{Niv09}.

\subsection{Deep Multiagent Reinforcement Learning}

Modern deep reinforcement learning methods take the perspective of an agent that must learn to maximize its cumulative long-term reward through trial-and-error interactions with its environment \cite{SuttonBarto:1998, Littman2015RL}.

In the multi-agent setting, the $i$-th agent stores a function $Q_i: \cO_i \times \cA_i \rightarrow \bR$ represented by a deep Q-network (DQN). See \cite{Mnih:2015} for details in the single agent case.
In our case the true state $s$ is observed differently by each player, as $o_i = O(s, i)$. However for consistency of notation, we use a shorthand: $Q_i(s,a) = Q_i(O(s,i), a)$.

During learning, to encourage exploration we parameterize the $i$-th agent's policy by
\begin{equation*}
 \pi_i(s) = \left\{ \begin{array}{ll}
                     \text{argmax}_{a \in \cA_i} Q_i(s, a) ~~ &\text{with probability} ~ 1 - \epsilon \\
                     \cU(\cA_i) ~~ &\text{with probability} ~ \epsilon \\
                    \end{array} \right.
\end{equation*}
where $\cU(\cA_i)$ denotes a sample from the uniform distribution over $\cA_i$.
Each agent updates its policy given a stored batch\footnote{The batch is sometimes called a ``replay buffer'' e.g.~\cite{Mnih:2015}.} of experienced transitions $\{(s, a, r_i, s')_t : t = 1, \dots T\}$ such that
\begin{equation*}
 Q_i(s, a) \gets Q_i(s, a) + \alpha \left[ r_i + \gamma \max_{a^\prime \in \cA_i}Q_i(s', a^\prime) - Q_i(s, a)\right]
\end{equation*}

This is a ``growing batch'' approach to reinforcement learning in the sense of~\cite{lange2012batch}. However, it does not grow in an unbounded fashion. Rather, old data is discarded so the batch can be constantly refreshed with new data reflecting more recent transitions. We compared batch sizes of 1e5 (our default) and 1e6 in our experiments (see Sect.~\ref{sec:comparison}).  The network representing the function $Q$ is trained through gradient descent on the mean squared Bellman residual with the expectation taken over transitions uniformly sampled from the batch (see~\cite{Mnih:2015}). Since the batch is constantly refreshed, the $Q$-network may adapt to the changing data distribution arising from the effects of learning on $\pi_1$ and $\pi_2$. 

In order to make learning in SSDs tractable, we make the extra assumption that each individual agent's learning depends only on the other agent's learning via the (slowly) changing distribution of experience it generates. That is, the two learning agents are ``independent'' of one another and each regard the other as part of the environment. From the perspective of player one, the learning of player two shows up as a non-stationary environment. The independence assumption can be seen as a particular kind of bounded rationality: agents do no recursive reasoning about one another's learning. In principle, this restriction could be dropped through the use of planning-based reinforcement learning methods like those of~\cite{kleimanWeiner2016}.

\section{Simulation Methods}\label{sec:methods}
Both games studied here were implemented in a 2D grid-world game engine. The state $s_t$ and the joint action of all players $\vec{a}$ determines the state at the next time-step $s_{t+1}$. Observations $O(s,i) \in \mathbb{R}^{3 \times 16 \times 21}$ (RGB) of the true state $s_t$ depended on the player's current position and orientation. The observation window extended 15 grid squares ahead and 10 grid squares from side to side (see Fig. \ref{fig:maps}B). Actions $a \in \mathbb{R}^8$ were agent-centered: step forward, step backward, step left, step right, rotate left, rotate right, use beam and stand still. Each player appears blue in its own local view, light-blue in its teammates view and red in its opponent's view. Each episode lasted for $1,000$ steps. Default neural networks had two hidden layers with 32 units, interleaved with rectified linear layers which projected to the output layer which had 8 units, one for each action. During training, players implemented epsilon-greedy policies, with epsilon decaying linearly over time (from 1.0 to 0.1). The default  per-time-step discount rate was $0.99$.

\begin{figure}[h]
\centering
\begin{minipage}{0.9\textwidth}
  \centering
  \includegraphics[width=\linewidth]{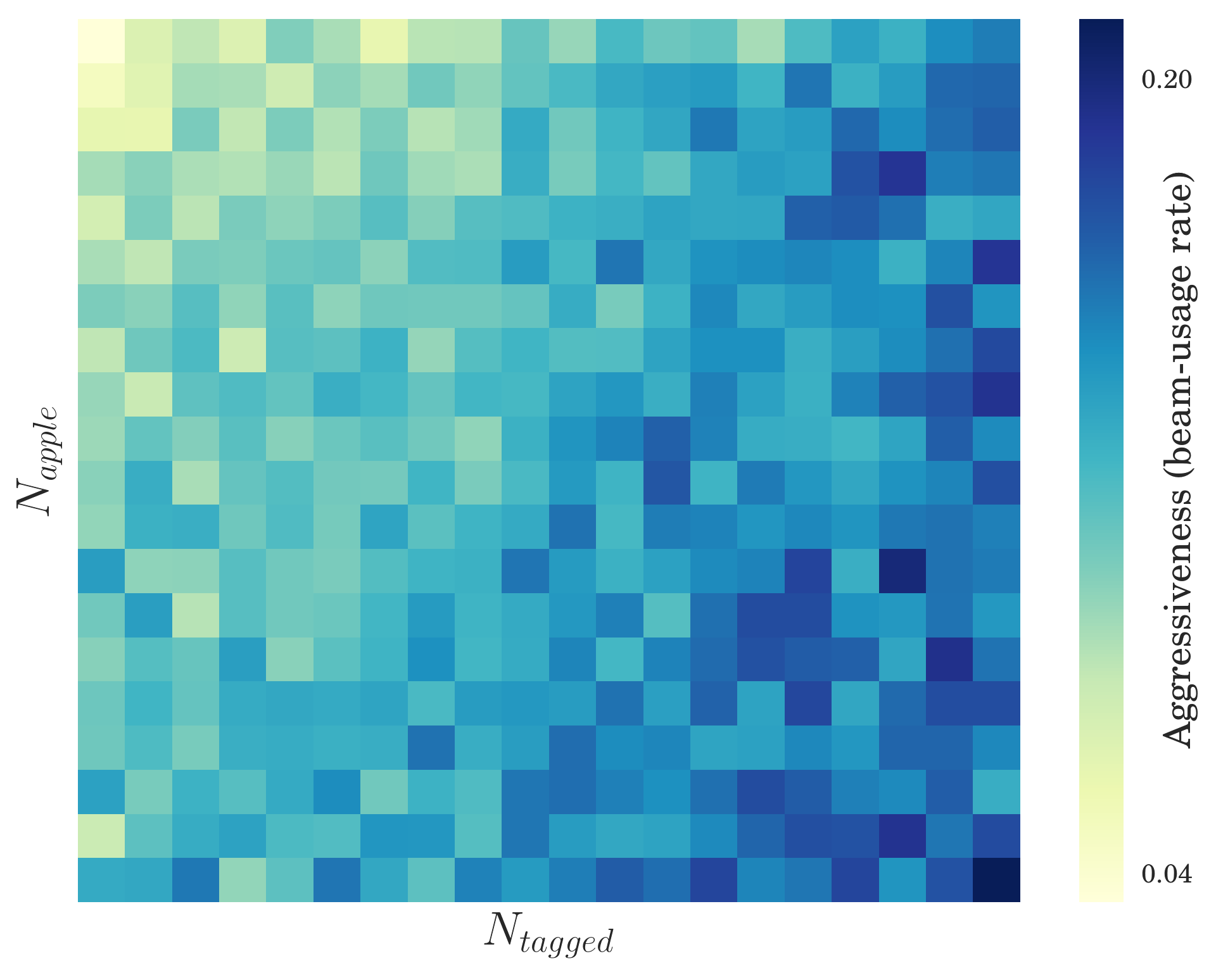}
  \includegraphics[width=\linewidth]{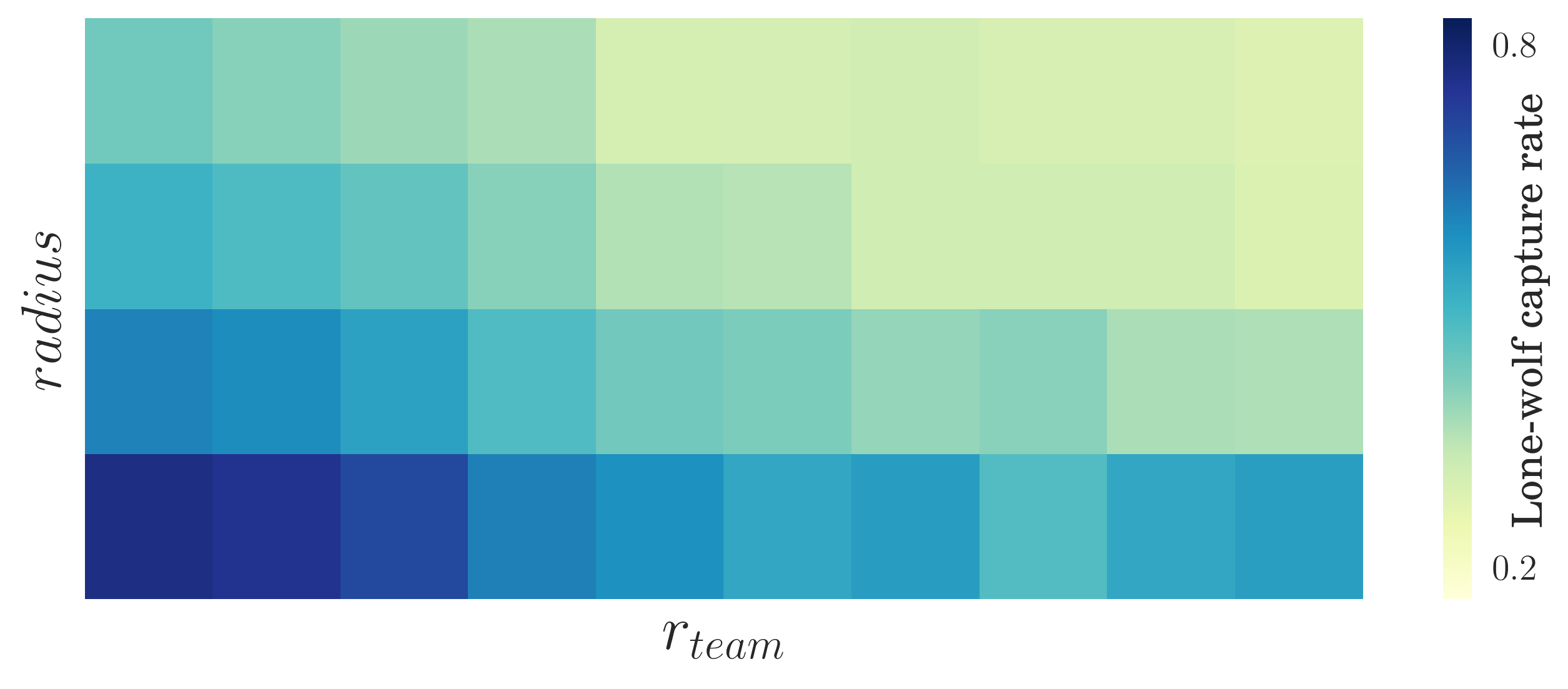}
\end{minipage}%
\caption{Social outcomes are influenced by environment parameters. Top: Gathering. Shown is the beam-use rate (aggressiveness) as a function of re-spawn time of apples $N_{\text{apple}}$ (abundance) and re-spawn time of agents $N_{\text{tagged}}$ (conflict-cost). These results show that agents learn aggresssive policies in environments that combine a scarcity of resources with the possibility of costly  action. Less aggressive policies emerge from learning in relatively abundant environments with less possibility for costly action. Bottom: Wolfpack. Shown is two minus the average number of wolves per capture as a function of the capture radius and group capture benefit $(r_{\text{team}})$. Again as expected, greater group benefit and larger capture radius lead to an increase in wolves per capture, indicating a higher degree of cooperation.} 
\label{fig:heatmaps}
\end{figure}

\section{Results}\label{sec:results}

In this section, we describe three experiments: one for each game (Gathering and Wolfpack),
and a third experiment investigating parameters that influence the emergence of cooperation versus defection.

\subsection{Experiment 1: Gathering}\label{sec:gathering}

The goal of the Gathering game is to collect apples, represented by green pixels (see Fig. \ref{fig:maps}A). When a player collects an apple it receives a reward of 1 and the apple is temporarily removed from the map. The apple respawns after $N_{\text{apple}}$ frames. Players can direct a beam in a straight line along their current orientation. A player hit by the beam twice is ``tagged'' and removed from the game for $N_{\text{tagged}}$ frames. No rewards are delivered to either player for tagging. The only potential motivation for tagging is competition over the apples. Refer to the Gathering gameplay video\footnote{\href{https://goo.gl/2xczLc}{https://goo.gl/2xczLc}} for demonstration.

Intuitively, a defecting policy in this game is one that is aggressive---i.e., involving frequent attempts to tag rival players to remove them from the game. Such a policy is motivated by the opportunity to take all the apples for oneself that arises after eliminating the other player. By contrast, a cooperative policy is one that does not seek to tag the other player. This suggests the use of a social behavior metric (section \ref{sec:define_ssd}) that measures a policy's tendency to use the beam action as the basis for its classification as defection or cooperation. To this end, we counted the number of beam actions during a time horizon and normalized it by the amount of time in which both agents were playing (not removed from the game).

\begin{figure*}[t]
\centering
\includegraphics[width=.9\linewidth]{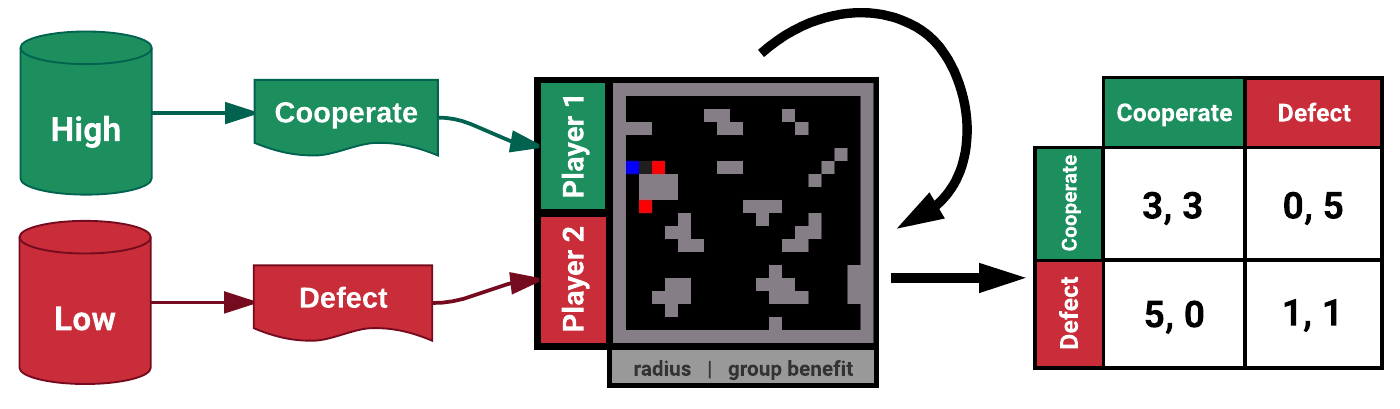}
\caption{Workflow to obtain empirical payoff matrices from Markov games. Agents are trained under different environmental conditions, e.g., with high or low abundance (Gathering case) or team capture bonus (Wolfpack case) resulting in agents classified as cooperators $(\pi^C \in \Pi^C)$ or defectors $(\pi^D \in \Pi^D)$. Empirical game payoffs are estimated by sampling $(\pi_1, \pi_2)$ from $\Pi^C \times \Pi^C$, $\Pi^C \times \Pi^D$, $\Pi^D \times \Pi^C$, and $\Pi^D \times \Pi^D$.  By repeatedly playing out the resulting games between the sampled $\pi_1$ and $\pi_2$, and averaging the results, it is possible to estimate the payoffs for each cell of the matrix.
\label{fig:tournament}}
\end{figure*}

By manipulating the rate at which apples respawn after being collected, $N_{\text{apple}}$, we could control the abundance of apples in the environment. Similarly, by manipulating the number of timesteps for which a tagged agent is removed from the game, $N_{\text{tagged}}$, we could control the cost of potential conflict. We wanted to test whether conflict would emerge from learning in environments where apples were scarce. We considered the effect of abundance $(N_{\text{apple}})$ and conflict-cost $(N_{\text{tagged}})$ on the level of aggressiveness (beam-use rate) that emerges from learning. Fig. \ref{fig:heatmaps}A shows the beam-use rate that evolved after training for 40 million steps as a function of abundance $(N_{\text{apple}})$ and conflict-cost $(N_{\text{tagged}})$.  Supplementary video \footnote{\href{https://goo.gl/w2VqlQ}{https://goo.gl/w2VqlQ}} shows how such emergent conflict evolves over the course of learning. In this case, differences in beam-use rate (proxy for the tendency to defect) learned in the different environments emerge quite early in training and mostly persist throughout. When learning does change  beam-use rate, it is almost always to increase it.

We noted that the policies learned in environments with low abundance or high conflict-cost were highly aggressive while the policies learned with high abundance or low conflict-cost were less aggressive. That is, the Gathering game predicts that conflict may emerge from competition for scarce resources, but is less likely to emerge when resources are plentiful.

\begin{figure*}[t]
\centering
\begin{minipage}{.45\textwidth}
  \centering
  \includegraphics[width=\linewidth]{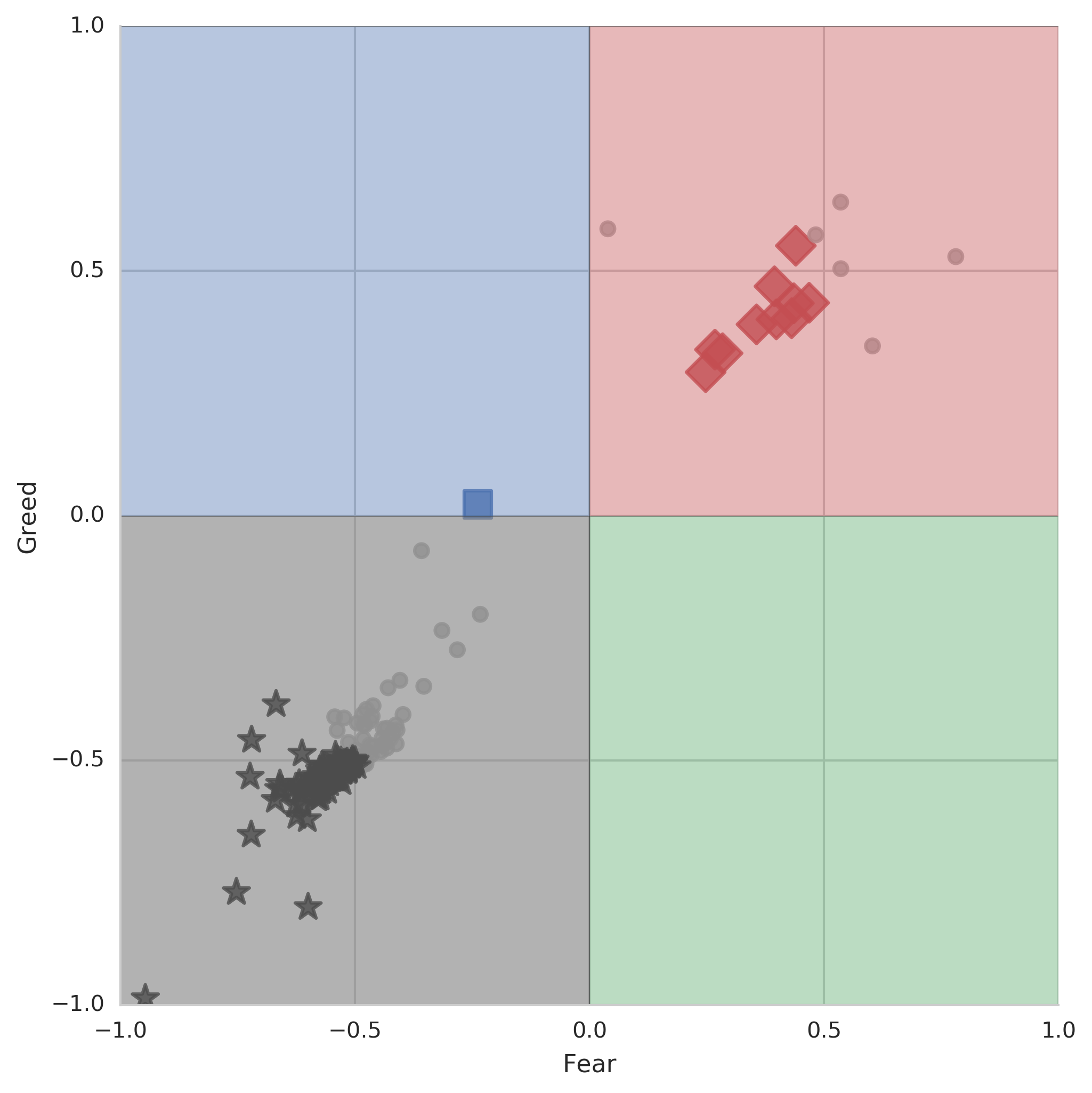}
\end{minipage}%
\begin{minipage}{.55\textwidth}
  \centering
  \includegraphics[width=\linewidth]{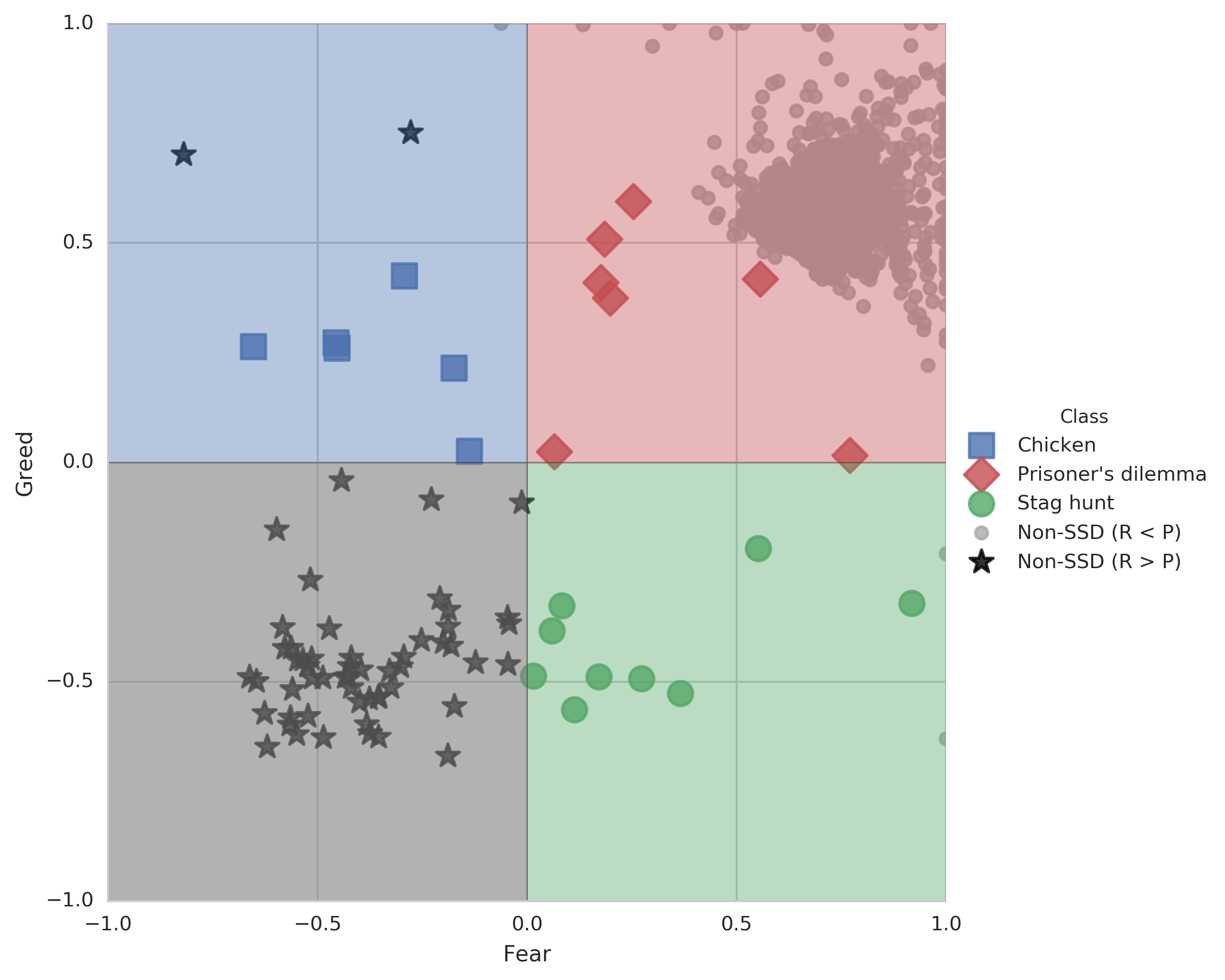}
\end{minipage}
\caption{Summary of matrix games discovered within Gathering (Left) and Wolfpack (Right) through extracting empirical payoff matrices. The games are classified by social dilemma type indicated by color and quandrant. With the x-axis representing $\text{fear} = P - S$ and the y-axis representing $\text{greed} = T - R$, the lower right quadrant contains Stag Hunt type games (green), the top left quadrant Chicken type games (blue), and the top right quadrant Prisoner's Dilemma type games (red). Non-SSD type games, which either violate social dilemma condition \eqref{eq:RgeP} or do not exhibit fear or greed are shown as well.
\label{fig:scatter_plots}}
\end{figure*}

To further characterize the mixed motivation structure of the Gathering game, we carried out the empirical game-theoretic analysis suggested by the definition of section \ref{sec:define_ssd}. We chose the set of policies $\Pi^C$ that were trained in the high abundance / low conflict-cost environments (low aggression policies) and $\Pi^D$ as policies trained in the low abundance and high conflict-cost  environments (high aggression policies), and used these to compute empirical payoff matrices as follows. Two pairs of policies $(\pi_1^C, \pi_1^D)$ and $(\pi_2^C, \pi_2^D)$ are sampled from $\Pi^C$ and $\Pi^D$ and matched against each other in the Gathering game for one episode. The resulting rewards are assigned to individual cells of a matrix game, in which $\pi_i^C$ corresponds the cooperative action for player $i$, and $\pi_j^D$, the defective action for player $j$. This process is repeated until convergence of the cell values, and generates estimates of $R,P,S,$ and $T$ for the game corresponding to each abundance / conflict-cost $(N_{\text{apple}}, N_{\text{tagged}})$ level tested. See Figure \ref{fig:tournament} for an illustration of this workflow. Fig. \ref{fig:scatter_plots}A summarizes the types of empirical games that were found given our parameter spectrum. Most cases where the social dilemma inequalities \eqref{eq:RgeP} -- \eqref{eq:greedOrfear} held, i.e., the strategic scenario was a social dilemma, turned out to be a prisoner's dilemma. The greed motivation reflects the temptation to take out a rival and collect all the apples oneself. The fear motivation reflected the danger of being taken out oneself by a defecting rival. $P$ is preferred to $S$ in the Gathering game because mutual defection typically leads to both players alternating tagging one another, so each gets some time alone to collect apples. Whereas the agent receiving the outcome $S$ does not try to tag its rival and thus never gets this chance.

\subsection{Experiment 2: Wolfpack}\label{sec:wolfpack}
The Wolfpack game requires two players (wolves) to chase a third player (the prey). When either wolf touches the prey, all wolves within the capture radius (see Fig. \ref{fig:maps}B) receive a reward. The reward received by the capturing wolves is proportional to the number of wolves in the capture radius. The idea is that a lone wolf can capture the prey, but is at risk of losing the carcass to scavengers. However, when the two wolves capture the prey together, they can better protect the carcass from scavengers and hence receive a higher reward. A lone-wolf capture provides a reward of $r_{\text{lone}}$ and a capture involving both wolves is worth $r_{\text{team}}$. Refer to the Wolfpack gameplay video\footnote{\href{https://goo.gl/AgXtTn}{https://goo.gl/AgXtTn}} for demonstration.

The wolves learn to catch the prey over the course of training. Fig. \ref{fig:heatmaps}B shows the effect on the average number of wolves per capture obtained from training in environments with varying levels of group capture bonus $r_{\text{team}}/r_{\text{lone}}$ and capture radius. Supplementary video \footnote{\href{https://goo.gl/vcB8mU}{https://goo.gl/vcB8mU}} shows how this dependency evolves over learning time. Like in the Gathering game, these results show that environment parameters influence how cooperative the learned policies will be. It is interesting that two different cooperative policies emerged from these experiments. On the one hand, the wolves could cooperate by first finding one another and then moving together to hunt the prey, while on the other hand, a wolf could first find the prey and then wait for the other wolf to arrive before capturing it.

Analogous to our analysis of the Gathering game, we choose $\Pi^C$ and $\Pi^D$ for Wolfpack to be the sets of policies learned in the high radius / group bonus and low radius /group bonus environments respectively. The procedure for estimating $R,P,S,$ and $T$ was the same as in section \ref{sec:gathering}. Fig. \ref{fig:scatter_plots}B summarizes these results. Interestingly, it turns out that all three classic MGSDs, chicken, stag hunt, and prisoner's dilemma can be found in the empirical payoff matrices of Wolfpack.

\subsection{Experiment 3: Agent parameters influencing the emergence of defection}\label{sec:comparison}
So far we have described how properties of the environment influence emergent social outcomes. Next we consider the impact of manipulating properties of the agents. Psychological research attempting to elucidate the motivational factors underlying human cooperation is relevant here. In particular, Social Psychology has advanced various hypotheses concerning psychological variables that may influence cooperation and give rise to the observed individual differences in human cooperative behavior in laboratory-based social dilemmas \cite{van2013psychology}. These factors include consideration-of-future-consequences \cite{kortenkamp2006time}, trust \cite{parks1995high}, affect (interestingly, it is \emph{negative} emotions that turn out to promote cooperation \cite{tan2010happiness}), and a personality variable called social value orientation characterized by other-regarding-preferences. The latter has been studied in a similar Markov game social dilemma setup to our SSD setting by \cite{Austerweil16}.

Obviously the relatively simple DQN learning agents we consider here do not have internal variables that directly  correspond to the factors identified by Social Psychology. Nor should they be expected to capture the full range of human individual differences in laboratory social dilemmas. Nevertheless, it is interesting to consider just how far one can go down this road of modeling Social Psychology hypotheses using such simple learning agents\footnote{The contrasting approach that seeks to build more structure into the reinforcement learning agents to enable more interpretable experimental manipulations is also interesting and complementary e.g., \cite{kleimanWeiner2016}.}. Recall also that DQN is in the class of reinforcement learning algorithms that is generally considered to be the leading  candidate theory of animal habit-learning \cite{daw2005uncertainty, Niv09}. Thus, the interpretation of our model is that it only addresses whatever part of cooperative behavior arises ``by habit'' as opposed to conscious deliberation.

\begin{figure*}[t]
  \centering
  \hspace*{-0.75cm}\includegraphics[width=1.05\linewidth]{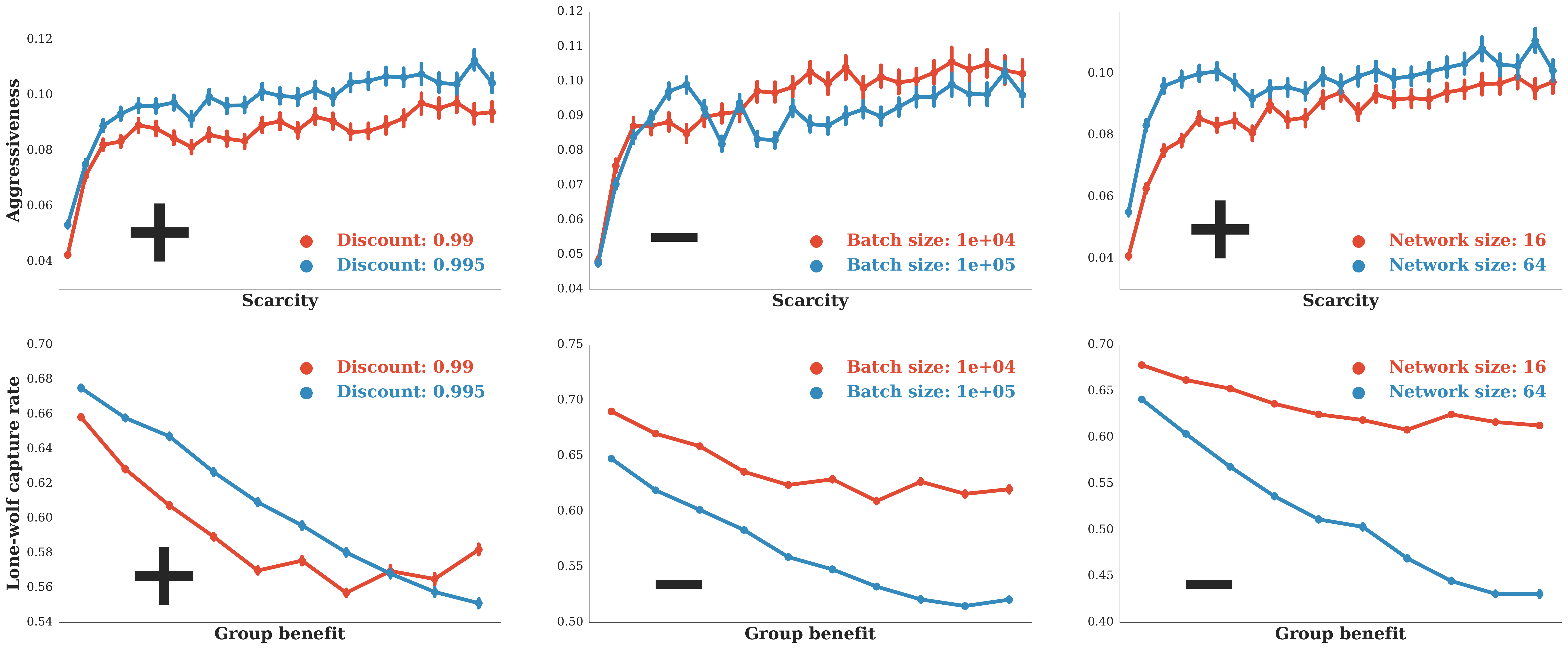}  
\caption{Factors influencing the emergence of defecting policies. Top row: Gathering. Shown are plots of average beam-use rate (aggressiveness) as a function of $N_{\text{apple}}$ (scarcity) Bottom row:  Wolfpack. Shown are plots of (two minus) average-wolves-per-capture (Lone-wolf capture rate) as a function of $r_{\text{team}}$ (Group Benefit). For both Gathering and Wolfpack we vary the following factors: temporal discount (left), batch size (centre), and network size (right). Note that the effects of discount factor and batch size on the tendency to defect point in the same direction for Gathering and Wolfpack, network size has the opposite effect (see text for discussion.)
\label{fig:headline_chart}}
\end{figure*}


Experimental manipulations of DQN parameters yield consistent and interpretable effects on emergent social behavior. Each plot in Fig.~\ref{fig:headline_chart} shows the relevant social behavior metric, conflict for Gathering and lone-wolf behavior for Wolfpack, as a function of an environment parameter: $N_{\text{apple}},N_{\text{tagged}}$ (Gathering) and $r_{\text{team}}/r_{\text{lone}}$ (Wolfpack). The figure shows that in both games, agents with greater discount parameter (less time discounting) more readily defect than agents that discount the future more steeply. For Gathering this likely occurs because the defection policy of tagging the other player to temporarily remove them from the game only provides a delayed reward in the form of the increased opportunity to collect apples without interference. However, when abundance is very high, even the agents with higher discount factors do not learn to defect. In such paradisiacal settings, the apples respawn so quickly that an individual agent cannot collect them quickly enough. As a consequence, there is no motivation to defect regardless of the temporal discount rate. Manipulating the size of the stored-and-constantly-refreshed batch of experience used to train each DQN agent has the opposite effect on the emergence of defection. Larger batch size translates into more experience with the other agent's policy. For Gathering, this means that avoiding being tagged becomes easier. Evasive action benefits more from extra experience than the ability to target the other agent. For Wolfpack, larger batch size allows greater opportunity to learn to coordinate to jointly catch the prey. 

Possibly the most interesting effect on behavior comes from the number of hidden units in the neural network behind the agents, which may be interpreted as their cognitive capacity. Curves for tendency to defect are shown in the right column of Fig.~\ref{fig:headline_chart}, comparing two different network sizes. For Gathering, an increase in network size leads to an increase in the agent's tendency to defect, whereas for Wolfpack the opposite is true: Greater network size leads to less defection. 

This can be explained as follows. In Gathering, defection behavior is more complex and requires a larger network size to learn than cooperative behavior. This is the case because defection requires the difficult task of targeting the opposing agent with the beam whereas peacefully collecting apples is almost independent of the opposing agent's behavior. In Wolfpack, cooperation behavior is more complex and requires a larger network size because the agents need to coordinate their hunting behaviors to collect the team reward whereas the lone-wolf behavior does not require coordination with the other agent and hence requires less network capacity. 

Note that the qualitative difference in effects for network size supports our argument that the richer framework of SSDs is needed to capture important aspects of real social dilemmas. This rather striking difference between Gathering and Wolfpack is invisible to the purely matrix game based MGSD-modeling. It only emerges when the different complexities of cooperative or defecting behaviors, and hence the difficulty of the corresponding learning problems is modeled in a sequential setup such as an SSD.


\section{Discussion}\label{sec:discussion}
In the Wolfpack game, learning a defecting lone-wolf policy is easier than learning a cooperative pack-hunting policy. This is because the former does not require actions to be conditioned on the presence of a partner within the capture radius. In the Gathering game the situation is reversed. Cooperative policies are easier to learn since they need only be concerned with apples and may not depend on the rival player's actions. However, optimally efficient cooperative policies may still require such coordination to prevent situations where both players simultaneously move on the same apple. Cooperation and defection demand differing levels of coordination for the two games. Wolfpack's cooperative policy requires greater coordination than its defecting policy. Gathering's defection policy requires greater coordination (to successfully aim at the rival player).

Both the Gathering and Wolfpack games contain embedded MGSDs with prisoner's dilemma-type payoffs. The MGSD  model thus regards them as structurally identical. Yet, viewed as SSDs, they make rather different predictions. This suggests a new dimension on which to investigate classic questions concerning the evolution of cooperation. For any to-be-modeled phenomenon, the question now arises: which SSD is a better description of the game being played? If Gathering is a better model, then we would expect cooperation to be the easier-to-learn ``default'' policy, probably requiring less coordination. For situations where Wolfpack is the better model, defection is the easier-to-learn ``default'' behavior and cooperation is the harder-to-learn policy requiring greater coordination. These modeling choices are somewhat orthogonal to the issue of assigning values to the various possible outcomes (the only degree of freedom in MGSD-modeling), yet they make a large difference to the results.

SSD models address similar research questions as MGSD models, e.g. the evolution of cooperation. However, SSD models are more realistic since they capture the sequential structure of real-world social dilemmas. Of course, in modeling, greater verisimilitude is not automatically virtuous. When choosing between two models of a given phenomenon, Occam's razor demands we prefer the simpler one. If SSDs were just more realistic models that led to the same conclusions as MGSDs then they would not be especially useful. This however, is not the case. We argue the implication of the results presented here is that standard evolutionary and learning-based approaches to modeling the trial and error process through which societies converge on equilibria of social dilemmas are unable to address the following important learning related phenomena.
\begin{enumerate}
 \item Learning which strategic decision to make, abstractly, whether to cooperate or defect, often occurs simultaneously with learning how to efficiently implement said decision.
 \item It may be difficult to learn how to implement an effective cooperation policy with a partner bent on defection---or vice versa.
 \item Implementing effective cooperation or defection may involve solving coordination subproblems, but there is no guarantee this would occur, or that cooperation and defection would rely on coordination to the same extent. In some strategic situations, cooperation may require coordination, e.g., standing aside to allow a partner's passage through a narrow corridor while in others defection may require coordination e.g. blocking a rival from passing.
 \item Some strategic situations may allow for multiple different implementations of cooperation, and each may require coordination to a greater or lesser extent. The same goes for multiple implementations of defection.
 \item The complexity of learning how to implement effective cooperation and defection policies may not be equal. One or the other might be significantly easier to learn---solely due to  implementation complexity---in a manner that cannot be accounted for by adjusting outcome values in an MGSD model.
\end{enumerate}

Our general method of tracking social behavior metrics in addition to reward while manipulating parameters of the learning environment is widely applicable. One could use these techniques to simulate the effects of external interventions on social equilibria in cases where the sequential structure of cooperation and defection are important. Notice that several of the examples in Schelling's seminal book \emph{Micromotives and Macrobehavior}~\cite{schelling1978micromotives} can be seen as temporally extended social dilemmas for which policies have been learned over the course of repeated interaction, including the famous opening example of lecture hall seating behavior. It is also possible to define SSDs that model the extraction of renewable vs  non-renewable resources and track the sustainability of the emergent social behaviors while taking into account the varying difficulties of learning sustainable (cooperating) vs. non-sustainable (defecting) policies. Effects stemming from the need to learn implementations for strategic decisions may be especially important for informed policy-making concerning such real-world social dilemmas.

\subsection*{Acknowledgments}
The authors would like to thank Chrisantha Fernando, Toby Ord, and Peter Sunehag for fruitful discussions in the lead-up to this work, and Charles Beattie, Denis Teplyashin, and Stig Petersen for software engineering support.

\bibliography{seqsocdim}

\begin{thebibliography}{10}

\bibitem{rapoport1974prisoner}
Anatol Rapoport.
\newblock Prisoner's dilemma--recollections and observations.
\newblock In {\em Game Theory as a Theory of a Conflict Resolution}, pages
  17--34. Springer, 1974.

\bibitem{van2013psychology}
Paul~AM Van~Lange, Jeff Joireman, Craig~D Parks, and Eric Van~Dijk.
\newblock The psychology of social dilemmas: A review.
\newblock {\em Organizational Behavior and Human Decision Processes},
  120(2):125--141, 2013.

\bibitem{macy2002learning}
Michael~W Macy and Andreas Flache.
\newblock Learning dynamics in social dilemmas.
\newblock {\em Proceedings of the National Academy of Sciences}, 99(suppl
  3):7229--7236, 2002.

\bibitem{trivers1971evolution}
Robert~L. Trivers.
\newblock The evolution of reciprocal altruism.
\newblock {\em Quarterly Review of Biology}, pages 35--57, 1971.

\bibitem{Axelrod84}
Robert Axelrod.
\newblock {\em The Evolution of Cooperation}.
\newblock Basic Books, 1984.

\bibitem{nowak1992tit}
Martin~A Nowak and Karl Sigmund.
\newblock Tit for tat in heterogeneous populations.
\newblock {\em Nature}, 355(6357):250--253, 1992.

\bibitem{nowak1993strategy}
Martin Nowak, Karl Sigmund, et~al.
\newblock A strategy of win-stay, lose-shift that outperforms tit-for-tat in
  the prisoner's dilemma game.
\newblock {\em Nature}, 364(6432):56--58, 1993.

\bibitem{nowak1998evolution}
Martin~A Nowak and Karl Sigmund.
\newblock Evolution of indirect reciprocity by image scoring.
\newblock {\em Nature}, 393(6685):573--577, 1998.

\bibitem{axelrod1986evolutionary}
Robert Axelrod.
\newblock An evolutionary approach to norms.
\newblock {\em American political science review}, 80(04):1095--1111, 1986.

\bibitem{mahmoud2016cooperation}
Samhar Mahmoud, Simon Miles, and Michael Luck.
\newblock Cooperation emergence under resource-constrained peer punishment.
\newblock In {\em Proceedings of the 2016 International Conference on
  Autonomous Agents \& Multiagent Systems}, pages 900--908. International
  Foundation for Autonomous Agents and Multiagent Systems, 2016.

\bibitem{sandholm96}
T.W. Sandholm and R.H. Crites.
\newblock Multiagent reinforcement learning in the iterated prisoner's dilemma.
\newblock {\em Biosystems}, 37(1--2):147--166, 1996.

\bibitem{MunozDeCote06}
Enrique~Munoz de~Cote, Alessandro Lazaric, and Marcello Restelli.
\newblock Learning to cooperate in multi-agent social dilemmas.
\newblock In {\em Proceedings of the Fifth International Joint Conference on
  Autonomous Agents and Multiagent Systems (AAMAS)}, 2006.

\bibitem{wunder10}
M.~Wunder, M.~Littman, and M.~Babes.
\newblock Classes of multiagent {Q}-learning dynamics with greedy exploration.
\newblock In {\em Proceedings of the 27th International Conference on Machine
  Learning}, 2010.

\bibitem{Zawadzki14}
Erik Zawadzki, Asher Lipson, and Kevin Leyton{-}Brown.
\newblock Empirically evaluating multiagent learning algorithms.
\newblock {\em CoRR}, abs/1401.8074, 2014.

\bibitem{Bloembergen15}
Daan Bloembergen, Karl Tuyls, Daniel Hennes, and Michael Kaisers.
\newblock Evolutionary dynamics of multi-agent learning: A survey.
\newblock {\em Journal of Artificial Intelligence Research}, 53:659--697, 2015.

\bibitem{nowak1992evolutionary}
Martin~A Nowak and Robert~M May.
\newblock Evolutionary games and spatial chaos.
\newblock {\em Nature}, 359(6398):826--829, 1992.

\bibitem{yu2015emotional}
Chao Yu, Minjie Zhang, Fenghui Ren, and Guozhen Tan.
\newblock Emotional multiagent reinforcement learning in spatial social
  dilemmas.
\newblock {\em IEEE Transactions on Neural Networks and Learning Systems},
  26(12):3083--3096, 2015.

\bibitem{ohtsuki2006simple}
Hisashi Ohtsuki, Christoph Hauert, Erez Lieberman, and Martin~A Nowak.
\newblock A simple rule for the evolution of cooperation on graphs and social
  networks.
\newblock {\em Nature}, 441(7092):502--505, 2006.

\bibitem{santos2006new}
Francisco~C Santos and Jorge~M Pacheco.
\newblock A new route to the evolution of cooperation.
\newblock {\em Journal of Evolutionary Biology}, 19(3):726--733, 2006.

\bibitem{walsh2002analyzing}
William~E Walsh, Rajarshi Das, Gerald Tesauro, and Jeffrey~O Kephart.
\newblock Analyzing complex strategic interactions in multi-agent systems.
\newblock In {\em AAAI-02 Workshop on Game-Theoretic and Decision-Theoretic
  Agents}, pages 109--118, 2002.

\bibitem{Wellman06}
Michael Wellman.
\newblock Methods for empirical game-theoretic analysis (extended abstract).
\newblock In {\em Proceedings of the National Conference on Artificial
  Intelligence (AAAI)}, pages 1552--1555, 2006.

\bibitem{Littman94markovgames}
M.~L. Littman.
\newblock Markov games as a framework for multi-agent reinforcement learning.
\newblock In {\em Proceedings of the 11th International Conference on Machine
  Learning (ICML)}, pages 157--163, 1994.

\bibitem{Nowe12}
Ann Now\'{e}, Peter Vrancx, and Yann-Micha\"{e}l~De Hauwere.
\newblock Game theory and multiagent reinforcement learning.
\newblock In Marco Wiering and Martijn van Otterlo, editors, {\em Reinforcement
  Learning: State-of-the-Art}, chapter~14. Springer, 2012.

\bibitem{kleimanWeiner2016}
Max Kleiman-Weiner, M~K Ho, J~L Austerweil, Michael~L Littman, and Josh~B
  Tenenbaum.
\newblock Coordinate to cooperate or compete: abstract goals and joint
  intentions in social interaction.
\newblock In {\em Proceedings of the 38th Annual Conference of the Cognitive
  Science Society}, 2016.

\bibitem{Mnih:2015}
V.~Mnih, K.~Kavukcuoglu, D.~Silver, A.~A. Rusu, J.~Veness, M.~G. Bellemare,
  A.~Graves, M.~Riedmiller, A.~K. Fidjeland, G.~Ostrovski, S.~Petersen,
  C.~Beattie, A.~Sadik, I.~Antonoglou, H.~King, D.~Kumaran, D.~Wierstra,
  S.~Legg, and D.~Hassabis.
\newblock Human-level control through deep reinforcement learning.
\newblock {\em Nature}, 518(7540):529--533, 2015.

\bibitem{Shoham07}
Y.~Shoham, R.~Powers, and T.~Grenager.
\newblock If multi-agent learning is the answer, what is the question?
\newblock {\em Artificial Intelligence}, 171(7):365--377, 2007.

\bibitem{Lagoudakis02}
M.~G. Lagoudakis and R.~Parr.
\newblock Value function approximation in zero-sum {M}arkov games.
\newblock In {\em Proceedings of the 18th Conference on Uncertainty in
  Artificial Intelligence (UAI)}, pages 283--292, 2002.

\bibitem{Perolat15}
J.~P\'{e}rolat, B.~Scherrer, B.~Piot, and O.~Pietquin.
\newblock Approximate dynamic programming for two-player zero-sum {M}arkov
  games.
\newblock In {\em Proceedings of the International Conference on Machine
  Learning (ICML)}, 2015.

\bibitem{Perolat16Softened}
J.~P\'{e}rolat, B.~Piot, M.~Geist, B.~Scherrer, and O.~Pietquin.
\newblock Softened approximate policy iteration for {M}arkov games.
\newblock In {\em Proceedings of the International Conference on Machine
  Learning (ICML)}, 2016.

\bibitem{Bosansky16Algorithms}
Branislav Bo\v{s}ansk\'{y}, Viliam Lis\'{y}, Marc Lanctot, Ji\v{r}\'{i}
  \v{C}erm\'{a}k, and Mark~H.M. Winands.
\newblock Algorithms for computing strategies in two-player simultaneous move
  games.
\newblock {\em Artificial Intelligence}, 237:1--40, 2016.

\bibitem{Zinkevich06}
M.~Zinkevich, A.~Greenwald, and M.~Littman.
\newblock Cyclic equilibria in {M}arkov games.
\newblock In {\em Neural Information Processing Systems}, 2006.

\bibitem{Hu98NashQ}
J.~Hu and M.~P. Wellman.
\newblock Multiagent reinforcement learning: Theoretical framework and an
  algorithm.
\newblock In {\em Proceedings of the 15th International Conference on Machine
  Learning (ICML)}, pages 242--250, 1998.

\bibitem{Greenwald03CEQ}
A.~Greenwald and K.~Hall.
\newblock Correlated-{Q} learning.
\newblock In {\em Proceedings of the 20th International Conference on Machine
  Learning (ICML)}, pages 242--249, 2003.

\bibitem{Littman01}
Michael Littman.
\newblock Friend-or-foe {Q}-learning in general-sum games.
\newblock In {\em Proceedings of the Eighteenth International Conference on
  Machine Learning}, pages 322--328, 2001.

\bibitem{Perolat16}
J.~P\'{e}rolat, B.~Piot, B.~Scherrer, and O.~Pietquin.
\newblock On the use of non-stationary strategies for solving two-player
  zero-sum {M}arkov games.
\newblock In {\em Proceedings of the 19th International Conference on
  Artificial Intelligence and Statistics}, 2016.

\bibitem{gmytrasiewicz2005framework}
Piotr~J Gmytrasiewicz and Prashant Doshi.
\newblock A framework for sequential planning in multi-agent settings.
\newblock {\em Journal of Artificial Intelligence Research}, 24:49--79, 2005.

\bibitem{varakantham2009exploiting}
Pradeep Varakantham, Jun-young Kwak, Matthew~E Taylor, Janusz Marecki, Paul
  Scerri, and Milind Tambe.
\newblock Exploiting coordination locales in distributed {POMDP}s via social
  model shaping.
\newblock In {\em Proceedings of the 19th International Conference on Automated
  Planning and Scheduling, ICAPS}, 2009.

\bibitem{becker2004solving}
Raphen Becker, Shlomo Zilberstein, Victor Lesser, and Claudia~V Goldman.
\newblock Solving transition independent decentralized {M}arkov decision
  processes.
\newblock {\em Journal of Artificial Intelligence Research}, 22:423--455, 2004.

\bibitem{Laurent11}
Guillaume~J. Laurent, La\"{e}titia Matignon, and N.~Le Fort-Piat.
\newblock The world of independent learners is not {M}arkovian.
\newblock {\em Int. J. Know.-Based Intell. Eng. Syst.}, 15(1):55--64, 2011.

\bibitem{Silver16Go}
David Silver, Aja Huang, Chris~J. Maddison, Arthur Guez, Laurent Sifre, George
  van~den Driessche, Julian Schrittwieser, Ioannis Antonoglou, Veda
  Panneershelvam, Marc Lanctot, Sander Dieleman, Dominik Grewe, John Nham, Nal
  Kalchbrenner, Ilya Sutskever, Timothy Lillicrap, Madeleine Leach, Koray
  Kavukcuoglu, Thore Graepel, and Demis Hassabis.
\newblock Mastering the game of {G}o with deep neural networks and tree search.
\newblock {\em Nature}, 529:484--489, 2016.

\bibitem{Schultz97}
W.~Schultz, P.~Dayan, and P.R. Montague.
\newblock A neural substrate of prediction and reward.
\newblock {\em Science}, 275(5306):1593--1599, 1997.

\bibitem{Niv09}
Y.~Niv.
\newblock Reinforcement learning in the brain.
\newblock {\em The Journal of Mathematical Psychology}, 53(3):139--154, 2009.

\bibitem{SuttonBarto:1998}
Richard~S. Sutton and Andrew~G. Barto.
\newblock {\em Introduction to Reinforcement Learning}.
\newblock MIT Press, 1998.

\bibitem{Littman2015RL}
Michael~L Littman.
\newblock Reinforcement learning improves behaviour from evaluative feedback.
\newblock {\em Nature}, 521(7553):445--451, 2015.

\bibitem{lange2012batch}
Sascha Lange, Thomas Gabel, and Martin Riedmiller.
\newblock Batch reinforcement learning.
\newblock In {\em Reinforcement learning}, pages 45--73. Springer, 2012.

\bibitem{kortenkamp2006time}
Katherine~V Kortenkamp and Colleen~F Moore.
\newblock Time, uncertainty, and individual differences in decisions to
  cooperate in resource dilemmas.
\newblock {\em Personality and Social Psychology Bulletin}, 32(5):603--615,
  2006.

\bibitem{parks1995high}
Craig~D Parks and Lorne~G Hulbert.
\newblock High and low trusters' responses to fear in a payoff matrix.
\newblock {\em Journal of Conflict Resolution}, 39(4):718--730, 1995.

\bibitem{tan2010happiness}
Hui~Bing Tan and Joseph~P Forgas.
\newblock When happiness makes us selfish, but sadness makes us fair: Affective
  influences on interpersonal strategies in the dictator game.
\newblock {\em Journal of Experimental Social Psychology}, 46(3):571--576,
  2010.

\bibitem{Austerweil16}
Joseph~L. Austerweil, Stephen Brawner, Amy Greenwald, Elizabeth Hilliard, Mark
  Ho, Michael~L. Littman, James MacGlashan, and Carl Trimbach.
\newblock How other-regarding preferences can promote cooperation in
  non-zero-sum grid games.
\newblock In {\em Proceedings of the AAAI Symposium on Challenges and
  Opportunities in Multiagent Learning for the Real World}, 2016.

\bibitem{daw2005uncertainty}
Nathaniel~D Daw, Yael Niv, and Peter Dayan.
\newblock Uncertainty-based competition between prefrontal and dorsolateral
  striatal systems for behavioral control.
\newblock {\em Nature neuroscience}, 8(12):1704--1711, 2005.

\bibitem{schelling1978micromotives}
Thomas~C. Schelling.
\newblock {\em Micromotives and macrobehavior}.
\newblock WW Norton \& Company, 1978 Rev. 2006.

\end{thebibliography}

\end{document}